\documentstyle[prl,aps,epsf]{revtex}
\begin{document}
\twocolumn[\hsize\textwidth\columnwidth\hsize \csname @twocolumnfalse\endcsname
\title{}
\title{
Magnetoresistance, Micromagnetism and Domain Wall Effects
in Epitaxial Fe and Co Structures with Stripe Domains (invited)}

\author{Andrew D. Kent*, Ulrich R\"{u}diger, Jun Yu}

\address{
Department of Physics, New York University,
4 Washington Place, New York, New York 10003, USA}

\author{Luc Thomas, Stuart S. P. Parkin}

\address{
IBM Research Division, Almaden Research Center,
San Jose, California  95120, USA}

\date{September 2, 1998}
 
\maketitle

\begin{abstract}
We review our recent magnetotransport and micromagnetic studies of 
lithographically defined epitaxial thin film structures of bcc 
Fe and hcp Co with stripe domains. Micromagnetic structure and 
resistivity anisotropy are shown to be the predominant sources 
of low field magnetoresistance (MR) in these microstructures,
with domain wall (DW) effects smaller but observable 
(DW-MR $\lesssim 1 \%$). In Fe, at low temperature, in a 
regime in which fields 
have a significant effect on electron trajectories, a novel 
negative DW contribution to the resistivity is observed. 
In hcp Co microstructures, temperature dependent transport 
measurements for current perpendicular and parallel
to walls show that any additional resistivity due to 
DW scattering is very small.  
\end{abstract}
\label{firstpage}
]
\narrowtext

\section{Introduction}
The effect of magnetic domain walls (DWs) on the magnetoresistance (MR) of
thin films, micro- and nanowires is a topic of great current interest. 
Recent experimental research has extended early work on 3d transition metal 
single crystals \cite{Coleman} to microfabricated structures of transition 
metals \cite{Viret,Gregg,Kent,Ruediger,RuedigerCo} and  transition metal 
alloys\cite{Shinjo,Dafine}.  
This topic has also been approached from a number of viewpoints.  In nanowires
an experimental goal has been to use MR to investigate DW
nucleation and dynamics in search of evidence for macroscopic quantum
phenomena \cite{Hong,Otani}. While in thin films and microstructures 
with stripe domains, experiments have
focused on understanding the basic mechanisms of DW scattering of 
conduction electrons \cite{Viret,Gregg,Kent,Ruediger,RuedigerCo}.
In both cases this experimental work has 
stimulated new theoretical work in this area, including 
studies of the effect of DWs on quantum transport in mesoscopic 
ferromagnets \cite{Tatara,Geller}.  Independently, a new mechanism 
of DW scattering was recently proposed which invokes the two channel 
model of conduction in ferromagnets and spin dependent electron 
scattering -- a starting point for understanding the phenomena of 
giant MR (GMR) \cite{Levy}. Another approach has extended
the two band model of Ref. \cite{Tatara} to general band structures and
state dependent scattering times, and obtained different results
\cite{Brataas}. 

In this article we review our recent experimental investigations of
patterned epitaxial thin film structures of Fe and Co with controlled 
stripe domains \cite{Kent,Ruediger,RuedigerCo}. 
Such materials have enabled detailed studies of the physical mechanisms 
by which the erasure of DWs and micromagnetic structure with applied 
magnetic fields produces MR. 
The MR phenomena observed in Fe and Co will be compared and contrasted. 
Experiments have revealed novel MR effects, 
including in the case of Fe at low temperature, a reduction in resistivity
when domains are present \cite{Ruediger} -- in contrast to the increase expected
due to DW scattering. This occurs in a low resistivity regime in which both the
influence of the internal field on electron trajectories and scattering of electrons
at film surfaces are important \cite{MML98}. 

While
at low temperature results on Fe and Co are different, 
at room temperature the MR 
behavior and the basic physical mechanisms of MR are the same.
For example, for fields applied parallel to the easy axis the MR is 
negative in both Fe and Co microstructures.  
In initial work on Co thin films with stripe 
domains this ``large'' negative MR was interpreted as evidence for the
newly proposed DW scattering mechanism, and a giant DW MR
\cite{Gregg}. Here MR measurements
as a function of the angle of the applied field and magnetic force microscopy 
(MFM) imaging in conjunction with micromagnetic simulations strongly 
suggest that this negative MR is mainly a conventional anisotropic MR (AMR)
effect -- and is thus a bulk scattering effect and {\em not} associated with 
DW scattering \cite{RuedigerCo}. Low temperature measurements on 
hcp Co microstructures 
as a function of the angle of the current and walls show that
any additional resistivity or MR associated with
DW scattering is very small.

\section{Fabrication and Magnetic Characterization}

For these studies microfabricated wires were prepared from
high quality bcc (110) Fe and hcp (0001) Co epitaxial thin films.  
These films were produced with an UHV e-beam evaporation system on a-axis
($11{\bar 2}$0) sapphire substrates using seeded epitaxial growth methods
\cite{Kent,Ruediger,RuedigerCo}.
A typical 100 nm thick Fe film prepared in this manner
had a residual resistivity ratio (RRR) of 30 and a low temperature resistivity of
$\rho_{o} = 0.2 \; \mu\Omega$cm.  
Co layers of thickness 55 nm, 70 nm, 145 nm and 185 nm (RRR= 19,
$\rho =0.16 \: \mu\Omega$cm) have been prepared and studied. 
The films were then patterned using projection optical lithography and ion-milling
to produce micron scale bars of 0.5 to 20 $\mu$m linewidth and $\sim$ 200 $\mu$m 
length for 4 point resistivity measurements. 

These films have a strong uniaxial component to the magnetic anisotropy.  For
(110) Fe thin films, shape anisotropy confines the magnetization to the
film plane which contains the easy [001], hard [$1{\bar 1}1$] and [$1{\bar 1}0$] 
intermediate 
magnetocrystalline axes. In contrast, hcp (0001) 
Co films have a strong uniaxial anisotropy with the magnetic easy axis 
perpendicular to the film plane.

A competition between the magnetocrystalline, exchange and magnetostatic 
interactions has been used to produce controlled stripe domain configurations in 
microfabricated structures. Fe films were patterned into wires with the 
long wire axis perpendicular to the [001] easy magnetic axis and
parallel to the [$1{\bar 1}0$] direction, which results in a pattern 
of regularly spaced stripe domains perpendicular to the long wire axis.  
Varying the linewidth changes the ratio of the magnetostatic and 
magnetocrystalline energies to the DW energy
and hence the domain size. Minimization of the free energy for such 
a situation leads to a simple scaling in which the domain width depends
on the square root of the wire linewidth \cite{Kittel}.
However, experiments reveal metastable domain configurations. Fig.~\ref{fig1}
shows magnetic force microscopy (MFM) images of microstructures of 
systematically varied linewidth in zero field performed at room temperature
with a vertically magnetized tip \cite{Knote2}. 
These images highlight the DWs and 
magnetic poles at the boundaries of the wires. Images in the left hand
column were taken after the wire had been saturated transverse to its long
axis, while those in the right hand column were taken after longitudinal magnetic
saturation. The domain size is seen to depend both on the 
linewidth and magnetic history.  The latter effect is particularly 
dramatic in the 2 $\mu$m wire (Fig.~\ref{fig1}a and b) where the domain width varies 
by a factor of 4, from 0.4 $\mu$m after longitudinal saturation to 1.6 
$\mu$m after transverse saturation. 
The domain width can be varied continuously in this range
by varying the angle of the in-plane saturating field prior to demagnetization 
\cite{RuedigerAPL}. Note also that the domain width
is considerably larger in 20 $\mu$m linewidth wires ($\sim 6 \: \mu $m Fig.~\ref{fig1}e and 1f). 

Domain configurations near the sample boundaries have an important influence
on the MR, and characterizing their influence is essential to the 
interpretation of these experiments.  Magnetic configurations at 
boundary surfaces
with normal vectors (${\bf \hat{n}}$) parallel to the magnetic easy axis
(${\bf \hat{e}}$) depend on the
ratio of the anisotropy to demagnetization energy, $Q=K/2\pi M^2_s$. 
For small $Q$ ($Q \ll 1$), flux closure domains 
(with ${\bf M} \perp  \bf{\hat{n}}$ to the boundary surface)
are favored to reduce the magnetostatic energy, while for large $Q$ 
($Q \gg 1 $) stripe domains which intersect the
surface with  ${\bf M} \parallel \bf{\hat{n}} \parallel \bf{\hat{e}}$ are
favored to reduce the magnetocrystalline energy density. 
Both Fe and Co are in the small $Q$ limit ($Q_{Fe}=0.03$ and $Q_{Co}=0.35$)
and flux closure domains are expected. For the Fe films these 
occur at the lithographically defined wire edges and for the Co films, 
which have perpendicular anisotropy, these form at the film top and bottom interfaces 
\cite{Hubert,Ebels}. 
Results and micromagnetic simulations for Co in this
geometry are discussed below. For Fe microstructures an approximate 
outline of the domain configuration 
is sketched in Fig.~\ref{fig1}a for a 2 $\mu$m linewidth wire 
after transverse saturation.

Since current is directed along the Fe wire, there are domains with magnetization {\bf M} 
oriented both parallel and perpendicular to the current density {\bf J}. In 
order to estimate the MR contributions due to resistivity anisotropy the 
volume fraction of closure domains (with {\bf M} $\parallel$ {\bf J}) has 
been estimated from MFM images and is labeled $\gamma$.  For the
2 $\mu$m linewidth this fraction is $\gamma= 0.4$ after longitudinal saturation and
$\gamma = 0.14$ after transverse saturation (see Ref. \cite{Ruediger}, Fig. 2).

\begin{figure}[bt]
\epsfxsize=2.95in
\vspace{7 cm}
\caption{MFM images in zero applied field of (a,b) 2 $\mu$m, (c,d) 5
$\mu$m, and (e,f) 20 $\mu$m linewidth Fe wires.  Images in the 
left hand column were taken after magnetic saturation transverse 
to the wire's long axis, while those in the right hand column were taken after 
longitudinal saturation. The dashed lines in (a) illustrates the flux 
closure domain configurations observed.}
\label{fig1}
\end{figure}
\noindent

\section{Magnetotransport Properties}
MR measurements were performed in a variable temperature
high field cryostat with the applied field oriented in  
three different orientations: (i) in-plane and perpendicular to
the wire long axis (transverse), (ii) in-plane and parallel
to the wire long axis (longitudinal), and (iii) perpendicular to the
film plane (perpendicular).  The sample was rotated between two of 
the three possible orientations in-situ (i.e., at the measurement temperature),
low excitation currents were used ($J < 10^4 A/cm^2$), 
and the magnetic history of the sample was carefully controlled.  

\subsection{Fe Microstructures}
\label{sec:Fe}

\begin{figure}[bt]
\epsfxsize=2.95in
\centerline{\epsfbox{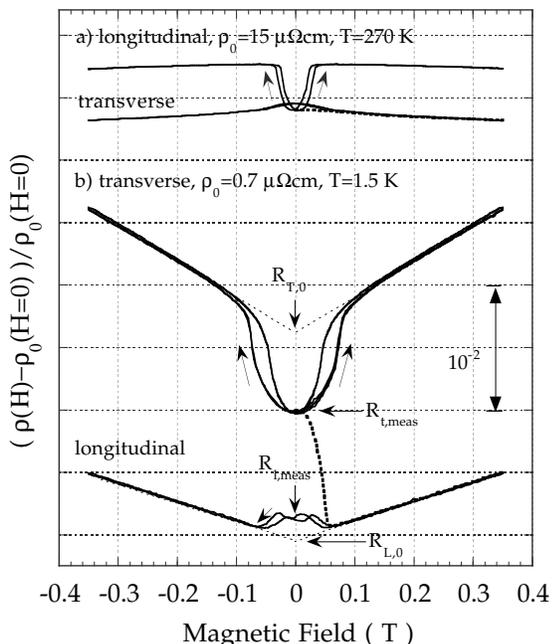}}
\caption{(a) MR data at $270$ K of a 2 $\mu$m linewidth Fe  wire in 
the transverse and a longitudinal field geometries.  
(b) MR at $1.5$ K again in the longitudinal and transverse 
field geometries.}
\label{fig2}
\end{figure}

Fig.~\ref{fig2} shows representative MR results on a 2 $\mu$m linewidth 
Fe wire for in-plane applied fields at both  a) high (270 K) and 
b) low temperature (1.5 K). First consider the
MR characteristics at 270 K. For 
fields transverse to the wire, and thus parallel to the magnetic
easy axis, the MR is negative. While in the longitudinal geometry,
the low field MR is positive. At fields greater
that the magnetic saturation fields ($H_{s\parallel}
= 0.035$ T and $H_{s\perp}= 0.085$T) the resistivity is essentially
independent of magnetic field and $\rho_{L}(H_s) > \rho_{T}(H_s)$.
At 1.5 K these characteristics change significantly.
The transverse MR is now positive, the MR is large and positive above
the saturation field and $\rho_{L}(H_s) < \rho_{T}(H_s)$, 
the resistivity anisotropy is reversed. 

\noindent 

\begin{figure}[bt]
\epsfxsize=2.95in
\centerline{\epsfbox{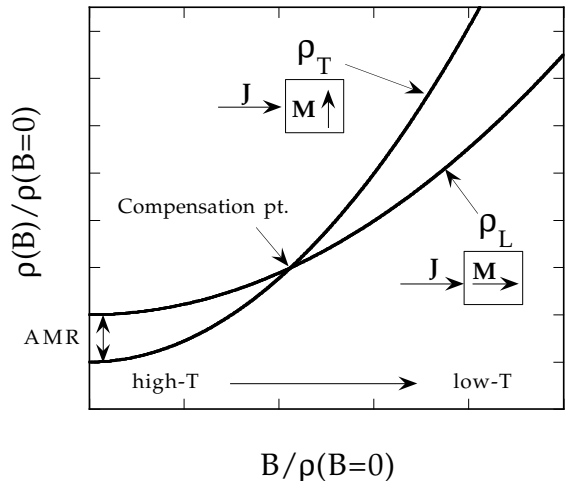}}
\caption{Schematic of the scaling of the MR and resistivity
anisotropy with field B, resistivity and temperature}
\label{fig3}
\end{figure}

This reversal is due to competing sources of resistivity anisotropy.  
The first is anisotropic MR (AMR) which has its origins in the spin-orbit
coupling -- the resistivity determined by extrapolation of MR data above
saturation to $B=0$ depends on the angle of $\bf{M}$ and $\bf{J}$ and, further, 
in a crystalline material, may depend on the direction of these vectors with
respect to the crystal axes \cite{McGuire}. Typically, AMR
in transition metals leads to $\rho_L(B=0)>\rho_T(B=0)$, 
that is domains with 
$\bf{M} \parallel \bf{J}$ have a greater resistivity than those with
$\bf{M} \perp \bf{J}$. However within magnetic domains, even in the
absence of externally applied fields, the $\bf{B}$ field can be large
(for Fe $4\pi M=2.2$ T and Co $4\pi M=1.8$ T, approximately
independent of temperature in the range studied)
and the 
anisotropy of the Lorentz MR can be important. The
Lorentz MR depends on the angle of $\bf{J}$ and $\bf{B}$ and is
a function of $B/\rho (B=0,T)$, the field divided by the zero field
resistivity, or equivalently, $\omega_c\tau$, the cyclotron frequency 
times the relaxation time.
Since the Lorentz force is proportional to  $\bf{J \times B}$, usually
$\rho_{T}(B) > \rho_{L}(B)$. At low temperature, due to the 
increase in the relaxation time, $\tau$, the anisotropy of the Lorentz
MR increases. As a result, at a certain temperature, the in-plane $H=0$ 
resistivity anisotropy changes sign \cite{Ruediger}. Fig.~\ref{fig3}
illustrates schematically this
scaling of the resistivity anisotropy with temperature and magnetic field. 
\noindent 

Resistivity anisotropy is a conventional source of low field MR. Starting 
from a multidomain sample, an applied saturating field
both erases DWs and reorients the magnetization with 
respect to the current direction and crystal axes. Since domains with $\bf{M} \perp \bf{J}$
and $\bf{M} \parallel \bf{J}$ have different resistivities this produces MR.
Quite distinct from DW contributions, this low field MR is associated 
with the electron scattering and orbital effects internal to domains 
discussed above.

This MR can be estimated within an effective medium model of the resistivity, 
assuming both that the domain size is greater than characteristic transport
lengths (such as the mean free path) and that the resistivity 
anisotropy is small.
The normalized resistivity measured at $H=0$ is given by 
\begin{equation}
R_{meas}(H=0) = \gamma R_{L,0} + (1 - \gamma) R_{T,0}
\label{eq:effres}
\end{equation}
Here $R_{L,0}$ and $R_{T,0}$ are the MRs extrapolated from above magnetic saturation
to $H=0$ (the dashed lines in Fig.~\ref{fig2}), and
$R_{meas}(H=0)$ is the measured normalized resistivity (as indicated in Fig.~\ref{fig2}). 
This simple model can 
account for the high temperature MR. For instance, the negative MR observed 
in the transverse 
geometry is due to erasure of higher resistivity closure domains (Fig.~\ref{fig2}a). Also, after
longitudinal saturation the volume of closure domains is smaller and thus the
resistivity at $H=0$ is lower.  Such a model has been employed to analyze the 
low temperature MR data as well \cite{Kent,Ruediger,Knote1}.

\begin{figure}[bt]
\epsfxsize=2.95in
\centerline{\epsfbox{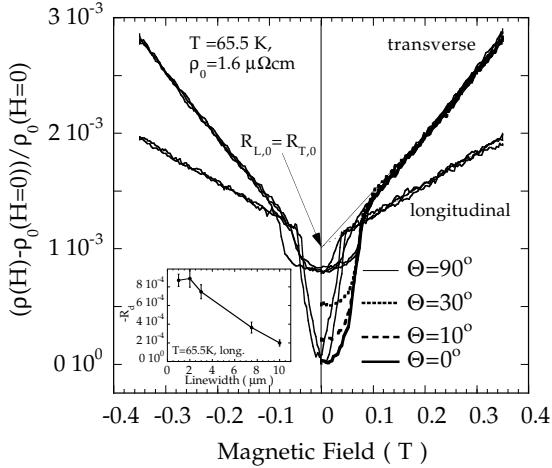}}
\caption{MR of a 2 $\mu$m linewidth Fe wire at $T_{comp}=65.5$ K. 
The extrapolation of the high field MR data to $H=0$ in transverse (solid 
line) and longitudinal (dashed line) geometry shows  $R_{L,0}= 
R_{T,0}$, the resistivity compensation at $H=0$. The resistivity with walls present, 
$\rho(H=0)$, is smaller than this extrapolation and indicates that the 
presence of DWs lowers the wire resistivity. The inset shows this 
negative DW contribution in the longitudinal geometry
as a function of linewidth at $T_{comp}$.}
\label{fig4}
\end{figure}
\noindent 

More directly, due to the competing contributions to the 
resistivity anisotropy, there
is a certain temperature at which the in-plane resistivity anisotropy vanishes.  
We denote this the compensation temperature $T_{comp}$ and it is defined as the 
temperature at which $R_{L,0}=R_{T,0}$. At this temperature the low field MR 
due to the in-plane reorientation of  wire magnetization should vanish.  This occurs
close to 65 K for the samples investigated.  Fig.~\ref{fig4} shows MR results at $T_{comp}$ 
for a 2 $\mu$m linewidth 100 nm thick Fe wire. 
The slope of the MR above the saturation field is due to the Lorentz effect and the 
extrapolation
of the high field MR to $H=0$ (dashed and solid lines) illustrates the resistivity
compensation. Fig.~\ref{fig4} shows that there remains a  positive low field MR,
in both the longitudinal ($\theta=0$) and transverse ($\theta=90$) 
field geometries, and the MR is 
greatest in the longitudinal geometry in which the
DW density at H=0 is largest (Fig.~\ref{fig1}b). Since the in-plane resistivity anisotropy 
is approximately zero at $T_{comp}$, these results have been 
taken as evidence for a negative DW 
contribution to the resistivity \cite{Ruediger}. The DW contribution to the 
MR is calculated at $T_{comp}$ as $R_d=R_{meas}(H=0)-R_{L,T,0}$ and is negative. 

By changing the angle of the demagnetizing field 
the density of DWs has been varied
continuously in a single sample between the limiting configurations seen in 
Fig.~\ref{fig1}a and b.  The magnitude
of the positive MR increases with increasing DW density (Fig.~\ref{fig4}) 
\cite{RuedigerAPL}. Varying the wire linewidth also varies the density of
DWs.  The inset of Fig.~\ref{fig4} shows that at $T_{comp}$ the magnitude of this negative DW 
contribution to the resistivity decreases with increasing wire linewidth and, hence,
decreasing DW density.

\begin{figure}[bt]
\epsfxsize=2.95in
\centerline{\epsfbox{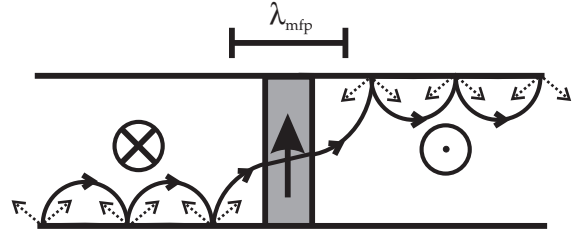}}
\caption{Cross-sectional view of the magnetic configuration of an Fe wire of
thickness $t$, showing the effect of internal fields and surface scattering
on the trajectory of charge carriers within stripe domains and DWs.}
\label{fig5}
\end{figure}
\noindent 

This effect has been studied further in microstructures of systematically 
varied film thickness \cite{MML98}.  In films of 200 nm thickness and greater
the negative DW contribution to the resistivity is reduced significantly.
These results have suggested a novel mechanism by which domains may increase 
conductivity in thin films. When diffuse electron scattering at the film 
top and bottom interfaces is important, as in the case of these high 
quality films at low temperature, the internal field
acting on electron trajectories near walls may act to deflect charge 
from the film interfaces and hence reduce resistivity (Fig.~\ref{fig5}). 
Increasing the film thickness acts to reduce the
importance of surface scattering and hence this effect.  

\subsection{Co Microstructures}
\label{sec:Co}

We now turn to transport studies of (0001) hcp Co microstructures. 
Fig.~\ref{fig6} shows MFM images of a 70 nm thick 5 $\mu$m linewidth Co 
wire in zero magnetic field. Images are shown after magnetic 
saturation: a) perpendicular to the film plane, 
b) in-plane and transverse to the wire axis, and c) in-plane 
and along the wire axis. An
in-plane applied field can be employed to align DWs in stripes
\cite{Kooy}. Fig.~\ref{fig6}b and c shows 
that DWs can be oriented parallel or perpendicular to the long axis of the
wire and thus the applied current, denoted as current-in-wall (CIW) 
and current-perpendicular-to-wall (CPW) geometries, respectively \cite{Levy} 
(as shown in the drawing in Fig.~\ref{fig6}). The magnetic structure of a cross-section 
of the film has been computed with the LLG Micromagnetics Simulator 
\cite{Scheinfein}. Details of this simulation can be found in Ref.
\cite{RuedigerCo}. Calculations for the film 
thicknesses studied give domain widths which are in excellent agreement
with experiment. For the 55 nm thick film a domain width of 66 nm was measured
with MFM and the calculated domain width was 64 nm. Fig.~\ref{fig6}d shows part of a
simulated magnetic cross-section of a 
55 nm thick Co element. The arrows indicate the magnetization direction
of the stripe and flux closure domains. In-plane magnetized
flux closure domains constitute approximately 33 \% of the 
total wire volume. Similarities between the computed magnetic structure 
of these films in cross-section 
and MFM images of Fe microstructures in the plane of the film are quite evident
(for example, compare this simulation to Fig.~\ref{fig1}a).

\begin{figure}[bt]
\epsfxsize=2.95in
\vspace{10 cm}
\caption{MFM images in zero applied field of a of 5 $\mu$m linewidth
70 nm thick Co wire after (a) perpendicular, (b) transverse, and 
(c) longitudinal magnetic saturation. The model shows the orientation
of stripe and flux closure domains with respect to the current for (b)
CPW and (c) CIW geometries. (d) A calculated magnetic domain cross-section 
of a 55 nm thick Co element showing out-of-plane magnetized stripe 
domains and in-plane magnetized flux closure domains.}
\label{fig6}
\end{figure}
\noindent 

\begin{figure}[bt]
\epsfxsize=2.95in
\centerline{\epsfbox{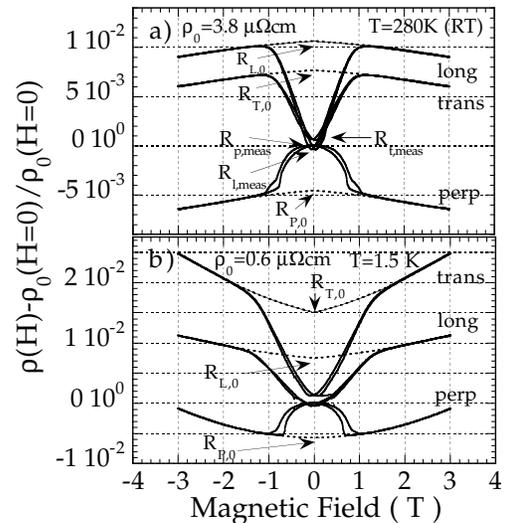}}
\caption{MR data of a 5 $\mu$m linewidth 55 nm thick Co wire in the 
perpendicular, transverse, and longitudinal field geometries at, (a) 270 K
and (b) 1.5 K.}
\label{fig7}
\end{figure}

The general features of the MR of these materials are also similar to those seen in Fe 
microstructures. Fig.~\ref{fig7} shows MR results for the 3 different field orientations at 
high (280 K) and low (1.5 K) temperatures.  
The low field MR is negative for fields
applied along the magnetic easy axis (i.e. perpendicular to the film plane), 
as in the case of Fe at high temperatures, and positive for in-plane applied fields 
(transverse and longitudinal geometry). At 280 K (Fig.~\ref{fig7}a), 
above the saturation field of 
$\sim 1.4$ T, 
there is a large anisotropy of the resistivity (and a small negative high field MR),
with the resistivity largest when the magnetization is in the film plane and
parallel to the current ($\bf{M} \parallel \bf{J}$).  As discussed above, this 
is typical of the resistivity anisotropy due to AMR.  
Note that the resistivity depends not only on the relative direction of 
$\bf{M}$ and $\bf{J}$, but also on the direction
that $\bf{M}$ makes with respect to the crystal axes, with the resistivity
smallest for $\bf{M} \perp \bf{J}$ and parallel to the [0001] direction. At low temperature
(1.5 K, Fig.~\ref{fig7}b) the resistivity is largest above the saturation field in the transverse
geometry, with $\bf{M} \perp \bf{J}$.  As in the case of Fe, and for the reasons already
discussed, the in-plane resistivity 
anisotropy changes sign with temperature. 
\noindent  

The resistivity anisotropy is again important in the interpretation
of the low field MR because the magnetization in zero applied field has
components in all three dimensions. For example, for the CPW geometry
(as illustrated in Fig.~\ref{fig6}), the magnetization
of the stripe domains are out-of-the-film plane and perpendicular to the 
current, the magnetization of 
the flux closure domains are in-plane and parallel to the current, and the 
magnetization of the Bloch wall rotates through an orientation in-plane and 
perpendicular to current. The low 
field MR which results from resistivity anisotropy and the reorientation of 
the film magnetization was neglected in the initial work on hcp Co films
\cite{Gregg}.

This contribution can be estimated again within an effective medium model 
of the resistivity. Starting from the maze configuration (Fig.~\ref{fig6}a) the 
perpendicular MR is:
\begin{equation}
R_{P,meas} - R_{P,0} = 
\gamma (\case{1}{2}(R_{L,0}+R_{T,0}) - R_{P,0})
\end{equation}
where $\gamma$ is the volume of in-plane magnetized closure domains.  Here 
$R_{L,T,P,0}$ are the MR extrapolated from high field
to H=0 (dashed lines in Fig.~\ref{fig7}) and $R_{P,meas}$ is the normalized resistivity
measured at 
H=0 in the maze configuration. In this expression, the small volume of
in-plane magnetized DW material has been neglected, only the flux closure
caps are considered.
From the MR measurements shown in Fig.~\ref{fig7}a
and with $\gamma = 0.33$, $R_{P,meas} - R_{P,0}$  is estimated to 
be $4.2 \: \times \: 10^{-3}$ at $280$ K, in close correspondence with the 
measured perpendicular MR.

The measured difference between CPW and CIW resistivities (i.e., 
$R_{t, meas} - R_{l, meas}$) in Fig.~\ref{fig6}a is given in terms of the resistivity
anisotropy as
\begin{equation}
R_{t,meas}-R_{l,meas}=\gamma (R_{L,0} - R_{T,0})
\end{equation}
which gives $ 1 \:  \times \: 10^{-3}$ at $280$ K, in close agreement with 
the experimental value.  
Although such estimates are qualitative (due to the uncertainties in 
the material magnetic structure and the applicability of such an effective 
medium model) they show that the predominate MR effects observed in this 
material are explicable by film micromagnetic structure and resistivity 
anisotropy, without the need to invoke DW scattering effects. Thus the simple 
MR measurements described cannot be used to unambiguously 
determine the intrinsic effect of DW scattering on resistivity.

\begin{figure}[bt]
\epsfxsize=2.95in
\centerline{\epsfbox{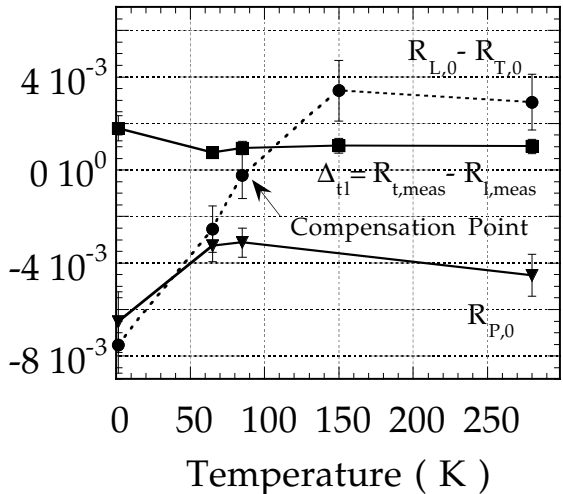}}
\caption{Measured resistivity anisotropy 
($R_{L,0} - R_{T,0}$), perpendicular MR ($R_{P,0}$) and 
difference between  CPW and CIP MRs ($\Delta_{tl}$) versus
temperature for a 5 $\mu$m linewidth 55 nm thick Co wire.}
\label{fig8}
\end{figure}
\noindent  

Temperature dependent resistivity measurements for CPW and CIW geometries
show more interesting behavior, which is not explicable simply in terms
of ferromagnetic resistivity anisotropy.
Since the in-plane resistivity anisotropy changes sign 
at low temperature, if the difference between CPW and CIW 
resistivities were only associated with resistivity anisotropy it would also
change sign on
lowering the temperature (Eqn. 3). Fig.~\ref{fig8} shows that this measured difference
$\Delta_{tl} \equiv R_{t,meas}-R_{l,meas}$ is always positive, while the 
resistivity anisotropy
($R_{L,0} - R_{T,0}$) is zero at $85 K$, the compensation temperature.
At $T_{comp}$,
$\Delta_{tl}$ is $ 9 \: \times \: 10^{-4}$.  This is consistent with 
a small additional resistivity due to DW scattering--since
we intuitively expect that this will lead to $R_{CPW}>R_{CIW}$.
If we assume that $\Delta_{tl}$ at $T_{comp}$ is due to DW scattering we
can estimate the order of magnitude of any putative DW scattering contribution
to the resistivity. Since walls will be much more effective at increasing
resistivity when arranged perpendicular to the current, we assume
DWs have only a small effect on resistivity
when parallel to the current \cite{Knote3}. 
The DW interface resistivity is then given
by $r= \frac{d}{\delta} \Delta_{tl} \rho_o \delta =
\Delta_{tl} \rho_o d$,
where d is the domain size, $\delta$ is the wall width ($\sim 15 \:  nm$)
and $\rho_o$ is the film resistivity. 
For the films
studied the average interface resistance is
$6 \: \pm \: 2 \: \times \: 10^{-19} \: \Omega m^2$ at $T_{comp}$ and the
MR due to the DW material,  
$\Delta \rho_{wall}/ \rho_o = \frac{d}{\delta} \Delta_{tl},$ is $0.5 \% $.
For comparison, these values are approximately a factor of 100 
smaller than the Co/Cu interface
resistance and MR in GMR multilayers with current 
perpendicular to the plane of the
layers \cite{Gijs}. 

\section{Summary}

In summary, these studies of epitaxial thin film structures
have revealed new MR phenomena and elucidated basic mechanism of MR in
microstructures with stripe domains. In both Fe and Co, micromagnetic
structure and resistivity anisotropy are the predominant sources of
low field MR. DWs have a smaller effect on MR.
In Fe at low temperatures DWs appear to enhance conductivity  via
the effect of internal fields on electron trajectories near DWs which act to
reduce scattering at film surfaces \cite{MML98}. In hcp Co, the
temperature dependence of the 
difference between CPW and CIW resistivities is evidence that any 
effects of DW interface scattering
on resistivity are quite small. 

These experiments, their analysis and interpretation suggest interesting
directions for future research.  
For example, with the recent theoretical interest in this area, 
they highlight the
necessity of experimentation on materials and nanofabricated structures
which will exemplify the transport physics associated with DWs. 
A further challenge is the extension of such research to nanometer scale
epitaxial structures
to explore, for example, quantum and ballistic transport phenomena 
in mesoscopic ferromagnets.

\section{Acknowledgments}

The authors thank Peter M. Levy for helpful discussions
of the work and comments on the manuscript. We thank M. Ofitserov
for technical assistance.
This research was supported by DARPA-ONR, Grant \# N00014-96-1-1207.  
Microstructures were prepared at the CNF, project \#588-96.  
\\
\\
$^*$Corresponding author: andy.kent@nyu.edu


\begin{references}
\bibitem{Knote2}
The fields from the MFM tip do not appear to affect the
domain structures, as the patterns observed do not change
during imaging.
\bibitem{Knote1}
The main uncertainty in
this analysis is
the dispersion of $\gamma$ found by MFM imaging (Fig.~\ref{fig1}), 
rather than in assumptions
about the wall structure within the film and the
wall thickness (30-50 nm), which we neglect.
\bibitem{Knote3}
For example, in the model of Ref. \cite{Levy}, a ratio of CPW to CIW MR due to
DW scattering is 7 (see their eqn. 21), for a spin assymmetry appropriate
for Co (ref. \cite{Gijs}) 
$\rho^{\uparrow}_o/\rho^{\downarrow}_o= 3$.
\end{references}
\end{document}